\documentclass[aps,prd,eqsecnum,12pt,showpacs,preprintnumbers,nofootinbib,%
superscriptaddress]{revtex4}
\usepackage{lscape}
\usepackage{indentfirst}
\usepackage{latexsym}
\usepackage{multirow}

\usepackage{amssymb,amsmath}
\usepackage{graphicx}
\newcommand{\bea}{\begin{eqnarray}}
\newcommand{\eea}{\end{eqnarray}}
\newcommand{\be}{\begin{equation}}
\newcommand{\ee}{\end{equation}}
\newcommand{\bes}{\begin{subequations}}
\newcommand{\ees}{\end{subequations}}

\begin{document}

\title{Noise kernel for a quantum field in Schwarzschild spacetime under
the Gaussian approximation}

\author{A. Eftekharzadeh}
\email{eftekhar@umd.edu}
\affiliation{Maryland Center for Fundamental Physics, Department of Physics,
University of Maryland, College Park, Maryland 20742-4111, USA}
\author{Jason D. Bates}
\email{batej6@wfu.edu}
\affiliation{Department of Physics, Wake Forest University,
Winston-Salem, North Carolina  27109, USA}
\author{Albert~Roura}
\email{albert.roura@aei.mpg.de}
\affiliation{Max-Planck-Institut f\"ur Gravitationsphysik
(Albert-Einstein-Institut), Am M\"uhlenberg 1, 14476 Golm, Germany}
\author{Paul R. Anderson}
\email{anderson@wfu.edu}
\affiliation{Department of Physics, Wake Forest University,
Winston-Salem, North Carolina  27109, USA}
\author{B. L. Hu}
\email{blhu@umd.edu}
\affiliation{Maryland Center for Fundamental Physics, Department of Physics,
University of Maryland, College Park, Maryland 20742-4111, USA}

\date{October 31, 2011}

\begin{abstract}

A method is given to compute an approximation to the noise kernel, defined as the symmetrized connected 2-point function of the stress tensor, for the conformally invariant scalar field in any spacetime conformal to an ultra-static spacetime for the case in which the field is in a thermal state at an arbitrary temperature.  The most useful applications of the method are flat space where the approximation is exact and Schwarzschild spacetime where the approximation is better than it is in most other spacetimes.  The two points are assumed to be separated in a timelike or spacelike direction.  The method involves the use of a Gaussian approximation which is of the same type as that used by Page~\cite{page} to compute an approximate form of the stress tensor for this field in Schwarzschild spacetime.  All components of the noise kernel have been computed exactly for hot flat space and one component is explicitly displayed.  Several components have also been computed for Schwarzschild spacetime and again one component is explicitly displayed.

\end{abstract}


\maketitle

\section{Introduction}

Studies of the fluctuations in the stress tensors of quantum fields are playing an increasingly important role in investigations of quantum effects in curved spacetimes  \cite{Ford82,Hu89,kuo-ford,wu-ford-1,wu-ford-2,PH97,PH00,PH00-2} and semiclassical gravity \cite{semiclassical gravity}. In various forms they have provided criteria for and tests of the validity of semiclassical gravity \cite{And-Mol-Mot-1,HRV04,perez-nadal08,And-Mol-Mot-2,froeb11b}. They are relevant for the generation of cosmological
perturbations during inflation~\cite{ford-structure-1,roura08,perez-nadal10,ford-structure-2,froeb11a} as well as the fluctuation and backreaction problem in black hole dynamics~\cite{HuRou07,HuRou07b}.
They also provide a possible pathway for one to connect semiclassical to quantum gravity~ (see, e.g., Ref.~\cite{HuKinTh}).  One important theory in which this is done systematically from first principles is stochastic semiclassical gravity
\cite{Hu99,martin99a,martin99b,martin00,stograCQG,stograLivRev}, or stochastic gravity in short,
which takes into account fluctuations of the gravitational field that are induced by the quantum matter fields.

In stochastic gravity the induced fluctuations of the gravitational field can be computed using the Einstein-Langevin equation~\cite{ELE,martin99a}
\be  G^{(1)}_{ab}[g+h] = 8 \pi (\langle \hat{T}^{(1)}_{ab}[g + h] \rangle + \xi_{ab}[g]) \;. \label{ein-lan-eq}  \ee
Here the superscript $(1)$ means that only terms linear in the metric perturbation $h_{ab}$ around the background geometry $g_{ab}$ should be kept, and
$g_{ab}$ is a solution to the semiclassical Einstein equation~\cite{semiclassical gravity,perez-nadal08}
\be G_{ab}[g] = 8 \pi \langle \hat{T}_{ab}[g] \rangle \;. \label{semiclass-eqs} \ee
Here $\langle \ldots \rangle$ denotes the quantum expectation value with respect to a normalized state of the matter field [more generally, $\langle \ldots \rangle = \mathrm{Tr}(\hat{\rho} \ldots)$ for a mixed state] and $\hat{T}_{ab}$ is the stress tensor operator of the field%
\footnote{The stress tensor expectation value in Eqs.~\eqref{ein-lan-eq}-\eqref{semiclass-eqs} is the renormalized one, which is the result of regularizing and subtracting the divergent terms by introducing appropriate local counterterms (up to quadratic order in the curvature) in the bare gravitational action \cite{birrell94}. Any finite contributions from those counterterms other than the Einstein tensor have been absorbed in the renormalized expectation value of the stress tensor.}.
The tensor $\xi_{ab}$ is a Gaussian stochastic source with vanishing mean which accounts for the stress tensor fluctuations and is completely characterized by its correlation function \cite{stograCQG,stograLivRev}:
\bea \langle \xi_{ab}(x)\rangle_s &=& 0 \nonumber \\
  \langle \xi_{ab}(x) \xi_{c'd'}(x')\rangle_s &=& N_{abc'd'}(x,x') \;,  \label{xi-properties}
\eea
where $\langle ... \rangle_s$ refers to the stochastic average over the realizations of the Gaussian source and the noise kernel $N_{abc'd'}(x,x')$ is given by the
the symmetrized connected 2-point function of the stress tensor operator for the quantum matter fields evaluated in the background geometry $g_{ab}$:
\bea N_{abc'd'} &=& \frac{1}{2} \langle \{\hat{t}_{ab}(x) ,\hat{t}_{c'd'}(x') \}\rangle     \label{NKdef}  \\
    \hat{t}_{ab}(x) &\equiv& \hat{T}_{ab}(x)- \langle \hat{T}_{ab}(x) \rangle  \;.
\eea
Thus, the noise kernel plays a central role in stochastic gravity, similarly to the expectation value of the stress tensor in semiclassical gravity.

One can solve Eq.~\eqref{ein-lan-eq} using the retarded Green function for the operator acting on $h_{ab}$ to obtain~\cite{HRV04}
\be h_{ab}(x) = h^\mathrm{(h)}_{ab}(x) + 8 \pi \! \int d^4 y' \sqrt{-g(y')}\,G^{\rm (ret)}_{abc'd'}(x,y')\, \xi^{c'd'}(y') \;, \label{h-sol}  \ee
where $h^\mathrm{(h)}_{ab}$ is a homogeneous solution to Eq.~\eqref{ein-lan-eq} which contains all the information on the initial conditions.
The resulting two-point function depends directly on the noise kernel:
\bea  && \!\!\!
\big\langle \langle h_{ab}(x) h_{c'd'}(x') \rangle_s \big\rangle_\mathrm{i.c.}
=  \langle h^\mathrm{(h)}_{ab}(x) h^\mathrm{(h)}_{c'd'}(x') \rangle_\mathrm{i.c.}
\nonumber \\
 && \; \; \; \; \;\; \; \; \; \;\; \; \; \; \;\; \; \; \; \;\; \; \; \;
 +\, (8 \pi)^2 \! \int d^4 y' d^4 y \sqrt{g(y') g(y)} \, G^{\rm (ret)}_{abe'f'}(x,y')\,
 N^{e' f' g h}(y',y)\, G^{\rm (ret)}_{c'd' g h}(x',y)  \,, \label{h-h-s}
 \; \; \; \;\; \; \; \;
\eea
where $\langle ... \rangle_\mathrm{i.c.}$ denotes the average over the initial conditions weighed by an appropriate distribution characterizing the initial quantum state of the metric perturbations.
It should be emphasized that although obtained by solving an equation involving classical stochastic processes, the result for the stochastic correlation function obtained in Eq.~\eqref{h-h-s} coincides with the result that would be obtained from a purely quantum field theoretical calculation where the metric is perturbatively quantized around the background $g_{ab}$. More precisely, if one considers a large number $N$ of identical fields, the stochastic correlation function coincides with the quantum correlation function $\langle \{\hat{h}_{ab}(x),\hat{h}_{c'd'}(x') \} \rangle$ to leading order in $1/N$ \cite{HRV04,CRVopensys}.
The noise kernel is the crucial ingredient in the contribution to the metric fluctuations \emph{induced} by the quantum fluctuations of the matter fields, which corresponds to the second term on the right-hand side of Eq.~\eqref{h-h-s}.

As pointed out by Hu and Roura~\cite{HuRou07} using the black-hole quantum backreaction and fluctuation problems as examples, a consistent study of the horizon fluctuations requires a detailed knowledge of the stress tensor 2-point function and, therefore, the noise kernel. That is because, in contrast with the averaged energy flux, the existence of a direct correlation assumed in earlier studies between the fluctuations of the energy flux crossing the horizon and those far from it is simply invalid.  The need for the noise kernel of a quantum field near a black hole horizon has been pronounced earlier in order to study the effect of
Hawking radiation emitted by a black hole on its evolution as well as the metric fluctuations driven by the quantum field (the ``backreaction and
fluctuation'' problem \cite{HRS}). For
example, Sinha, Raval and Hu \cite{SRH} have outlined a program for
such a study, which is the stochastic gravity upgrade (via the Einstein-Langevin equation)
of those carried out for the mean field in semiclassical gravity (through the semiclassical
Einstein equation) by York~\cite{york85,York} and by York and his collaborators~\cite{York2}.
Note that strictly speaking the retarded propagator and the noise kernel in Eq.~\eqref{h-h-s} should not be computed in the Schwarzschild background but a slightly corrected one (still static and spherically symmetric) which takes into account the backreaction of the quantum matter fields on the mean geometry via the semiclassical Einstein equation \cite{york85}. However, one can consider an expansion in powers of $1/M^2$; for the Hartle-Hawking state the difference between calculations employing the Schwarzschild background or the semiclassically corrected one would be of order $1/M^2$ or higher. Since our approach, which fits naturally within the framework of perturbative quantum gravity regarded as a low-energy effective theory \cite{burgess04}, is only valid for black holes with a Schwarzschild radius much larger than the Planck length ($M \gg 1$), those corrections of order $1/M^2$ will be very small.

An expression for the noise kernel for free fields in a general curved spacetime in terms of the corresponding Wightman function was obtained a decade ago \cite{martin99b,roura99b,PH01}. Since then this general result has been employed to obtain the noise kernel in Minkowski~\cite{martin00}, de Sitter~\cite{roura99b,perez-nadal10,perez-nadal11,park11} and anti-de Sitter~\cite{perez-nadal11,ChoHu noise kernelAdS}
spacetimes, as well as hot flat space and Schwarzschild spacetime in the coincidence limit~\cite{PH03}.
In this paper we compute expressions for the noise kernel in hot flat space and Schwarzschild spacetime using the same Gaussian approximation for the Wightman function of the quantum matter field that was used in Ref.~\cite{PH03}.  The difference is that there the coincident limit was considered and certain terms had to be subtracted. Here we do not take the coincidence limit and no subtraction of divergent terms is necessary.
In contrast to the stress tensor expectation value, which is computed in the
limit that the points come together, if one wishes to solve the equations of
stochastic semiclassical gravity, it is necessary to have an expression for the
noise kernel when the points are separated. This can be seen explicitly in
Eqs.~\eqref{xi-properties} and~\eqref{h-h-s}.

Specifically, we calculate an exact expression for the noise kernel of a conformally invariant scalar field in Minkowski space in a thermal state at an arbitrary temperature $T$.   We also compute approximate expressions for the noise kernel in both the optical Schwarzschild and Schwarzschild spacetimes when the field is in a thermal state at an arbitrary temperature $T$.  For the latter case when the temperature is that associated with the black hole, the field is in the Hartle-Hawking state, which is the relevant one if one wants to study the metric fluctuations of a black hole in (possibly unstable) equilibrium. In all cases the calculations are done with the points separated (and non-null related).  Because we do not attempt to take the limit in which the points come together (or are null related) the results are finite without the need for any subtraction.

To compute the noise kernel we need an expression for the Wightman function, $G^+(x,x') = \langle \phi(x) \phi(x') \rangle$.  We begin by working in the Euclidean sector of the optical Schwarzschild spacetime, which is ultra-static and conformal to Schwarzschild.   We use the same Gaussian approximation for the Euclidean Green function when the field is in a thermal state that Page~\cite{page} used for his computation of the stress tensor in Schwarzschild spacetime.  As he points out, this approximation corresponds to taking the first term in the DeWitt-Schwinger expansion for the Euclidean Green function.  In most spacetimes that would not be sufficient to generate an approximation to the stress tensor which could be renormalized correctly.
However, in the optical Schwarzschild spacetime (and for any other ultra-static
metric conformal to an Einstein metric in general) the second and third terms in
the DeWitt-Schwinger expansion vanish identically, so that the approximation is
much better than it would usually be.  In the flat space limit this expression
is exact.

Having obtained an expression for the Euclidean Green function, we analytically continue it to the Lorentzian sector and take the real part of the result to obtain an expression for the Wightman function when the points are non-null separated.  In the optical Schwarzschild spacetime, by substitution into the equation satisfied by
the Wightman function, we show that it is valid through $O[(x-x')^2]$ as expected%
\footnote{The dimensionless quantity which should be much smaller than one so that this expansion provides a valid approximation corresponds to the square of the geodesic distance, $2\sigma = O[(x-x')^2]$, divided by the square of the typical curvature radius scale. In Schwarzschild spacetime the latter is characterized near the horizon by the Schwarzschild radius $R_S = 2M$, but more accurately by the inverse fourth root of the curvature invariant $R_{\mu\nu\rho\sigma} R^{\mu\nu\rho\sigma} \sim M^2 / r^6\,$ far from it.%
\label{foot:expansion}}.
This expression is again exact in the flat space limit.

For hot flat space, we next compute the necessary derivatives of the Wightman function and substitute into Eq.~\eqref{general-noise-kernel} to obtain an exact expression for the noise kernel.  For the optical Schwarzschild
spacetime we take a different approach.  We expand the approximate part of the Wightman function in powers of $(x-x')$, compute the derivatives and substitute the results into Eq.~\eqref{general-noise-kernel}.
The result is valid through $O[(x-x')^{-4}]$, while the leading terms are $O[(x-x')^{-8}]$.  Finally, we conformally transform the results to Schwarzschild spacetime to obtain an expression for the noise kernel that is valid to the same order there.  This is done explicitly for two cases of interest.  One is the case when the point separation is only in the time direction and the product of the temperature and the point separation is not assumed to be small.  The second is for a general spacelike or timelike separation of the points when the product of the point separation and the temperature is assumed to be small.

All nonzero components of the noise kernel have been computed in hot flat space
for a non-null separation of the points and the conservation laws given in
Eq.~\eqref{conservation} and the partial traces given in
Eq.~\eqref{partial-traces} have been checked. Several components of the noise
kernel have also been computed in Schwarzschild spacetime, but for the sake of
brevity only one component is explicitly displayed.
We have obtained results for separations along the time direction but without
assuming the product of the temperature and the time difference to be small, as
well as for arbitrary separations but assuming that the product of the
temperature and the separation is small. In all cases, the relevant
conservation laws and partial traces have been checked. Furthermore, the result
which is not restricted to small values of the temperature times the time
difference is shown to agree with previously computed results in two different
flat space limits.

In Sec.~\ref{sec:general} we review the form of the noise kernel for a conformally invariant scalar field in a general spacetime and discuss its properties including its change under conformal transformations, which enables us to obtain the noise kernel in Schwarzschild spacetime from the result for the optical spacetime.
In Sec.~\ref{sec:gaussian} we present the relationship between the Wightman and Euclidean Green functions, the relevant parts of the formalism for
the DeWitt-Schwinger expansion~\cite{christensen}, and its use in the
Gaussian approximation derived by Page \cite{page} for the Euclidean Green function in the optical Schwarzschild spacetime.  We show that the resulting expression for the Wightman function is valid through $O[(x-x')^2]$ and for any temperature.
In Sec.~\ref{sec:computation-nk} a method for obtaining the Wightman function in the Gaussian approximation is given.
The computation of the noise kernel using this Wightman function is described and
 one component of the noise kernel in Schwarzschild spacetime is explicitly
displayed. The flat space limit for this component is discussed.
Sec.~\ref{sec:discussion} contains a summary and discussion of our main results.  In the Appendix we provide two proofs of the simple rescaling under conformal transformations of the noise kernel for a conformally invariant scalar field.
Throughout we use units such that $\hbar = c = G = k_B = 1$ and the sign conventions of Misner, Thorne, and Wheeler~\cite{MTW}.


\section{Noise kernel for the conformally invariant scalar fields}
\label{sec:general}

In this section we review the general properties of the noise kernel for the conformally invariant scalar field in an arbitrary
spacetime. The definition of the noise kernel for any quantized matter field is given in Eq.~\eqref{NKdef}.

The classical stress tensor for the conformally invariant scalar field is
\begin{equation}
T_{ab} = \nabla_a \phi \nabla_b \phi
- \frac{1}{2} g_{ab} \nabla^c \phi \nabla_c \phi
+ \xi \left( g_{ab} \Box - \nabla_a \nabla_b + G_{ab}\right) \phi^2 ,  \label{Tab}
\end{equation}
with $\xi =(D-2)/4(D-1)$, which becomes $\xi =1/6$ in $D=4$ spacetime dimensions.
Note that since the stress tensor is symmetric, the noise kernel as defined in Eq.~\eqref{NKdef} is also symmetric under exchange of the indices $a$ and $b$, or $c\,'$ and $d\,'$. To compute the noise kernel one promotes the field $\phi(x)$ in Eq.~\eqref{Tab} to an operator in the Heisenberg picture while treating $g_{ab}$ as a classical background metric.  The result is then substituted into Eq.~\eqref{NKdef}.

Given a Gaussian state for the quantum matter field, one can express the noise kernel in terms of products of two Wightman functions by applying Wick's theorem.\footnote{
Gaussian states include for instance the usual vacua, thermal and coherent states. In general most states will not be Gaussian and Wick's theorem will not apply: it will not be possible to write the 4-point function of the field in terms of 2-point functions and expectation values of the field. That is for example the case for all eigenstates of the particle number operators other than the corresponding vacuum.}
The Wightman function is defined as
\be G^+(x,x') = \langle \phi(x) \phi(x') \rangle \;.  \label{Wightman-def}  \ee
The result for a scalar field with arbitrary mass and curvature coupling in a general spacetime
has been obtained in Refs.~\cite{martin99b,PH01}.  For the conformally invariant scalar field in a general spacetime the noise
kernel is~\cite{PH01}
%
%
%
\begin{equation}
 N_{abc'd'}  =
 {\rm Re} \left\{  \bar K_{abc'd'}
  + g_{ab}   \bar K_{c'd'}
 + g_{c'd'} \bar K'_{ab}
 + g_{ab}g_{c'd'} \bar K \right\}
\label{general-noise-kernel}
\end{equation}
with\footnote{Notice that these equations have two slight but
crucial differences with the equations of Ref.~\cite{PH01}.
The sign of the last term of the equation for $N_{abc'd'}$ and also
the sign of the term $G\, G_{;abp'}^{\ \ \ \ p'}$ have been corrected.}
\bes
\label{General-Noise}
\begin{eqnarray}
9  \bar K_{abc'd'} &=&
%
4\,\left( G{}\!\,_{;}{}_{c'}{}_{b}\,G{}\!\,_{;}{}_{d'}{}_{a} +
    G{}\!\,_{;}{}_{c'}{}_{a}\,G{}\!\,_{;}{}_{d'}{}_{b} \right)
%
+ G{}\!\,_{;}{}_{c'}{}_{d'}\,G{}\!\,_{;}{}_{a}{}_{b} +
  G\,G{}\!\,_{;}{}_{a}{}_{b}{}_{c'}{}_{d'} \cr
%
&& -2\,\left( G{}\!\,_{;}{}_{b}\,G{}\!\,_{;}{}_{c'}{}_{a}{}_{d'} +
    G{}\!\,_{;}{}_{a}\,G{}\!\,_{;}{}_{c'}{}_{b}{}_{d'} +
    G{}\!\,_{;}{}_{d'}\,G{}\!\,_{;}{}_{a}{}_{b}{}_{c'} +
    G{}\!\,_{;}{}_{c'}\,G{}\!\,_{;}{}_{a}{}_{b}{}_{d'} \right)  \cr
%
&& + 2\,\left(
G{}\!\,_{;}{}_{a}\,G{}\!\,_{;}{}_{b}\,{R{}_{c'}{}_{d'}} +
    G{}\!\,_{;}{}_{c'}\,G{}\!\,_{;}{}_{d'}\,{R{}_{a}{}_{b}} \right)  \cr
%
&& - \left( G{}\!\,_{;}{}_{a}{}_{b}\,{R{}_{c'}{}_{d'}} +
  G{}\!\,_{;}{}_{c'}{}_{d'}\,{R{}_{a}{}_{b}}\right) G
%
 +{\frac{1}{2}}  {R{}_{c'}{}_{d'}}\,{R{}_{a}{}_{b}} {G^2}
\end{eqnarray}
\begin{eqnarray}
 36  \bar K'_{ab} &=&
%
8 \left(
 -  G{}\!\,_{;}{}_{p'}{}_{b}\,G{}\!\,_{;}{}^{p'}{}_{a}
 + G{}\!\,_{;}{}_{b}\,G{}\!\,_{;}{}_{p'}{}_{a}{}^{p'} +
  G{}\!\,_{;}{}_{a}\,G{}\!\,_{;}{}_{p'}{}_{b}{}^{p'}
\right)\cr &&
%
4 \left(
    G{}\!\,_{;}{}^{p'}\,G{}\!\,_{;}{}_{a}{}_{b}{}_{p'}
  - G{}\!\,_{;}{}_{p'}{}^{p'}\,G{}\!\,_{;}{}_{a}{}_{b} -
  G\,G{}\!\,_{;}{}_{a}{}_{b}{}_{p'}{}^{p'}
\right) \cr
%
&& - 2\,{R'}\,\left( 2\,G{}\!\,_{;}{}_{a}\,G{}\!\,_{;}{}_{b} -
    G\,G{}\!\,_{;}{}_{a}{}_{b} \right)  \cr
%
&&  -2\,\left( G{}\!\,_{;}{}_{p'}\,G{}\!\,_{;}{}^{p'} -
    2\,G\,G{}\!\,_{;}{}_{p'}{}^{p'} \right) \,{R{}_{a}{}_{b}}
%
 - {R'}\,{R{}_{a}{}_{b}} {G^2}
\end{eqnarray}
\begin{eqnarray}
 36 \bar K &=&
2\,G{}\!\,_{;}{}_{p'}{}_{q}\,G{}\!\,_{;}{}^{p'}{}^{q}
+ 4\,\left( G{}\!\,_{;}{}_{p'}{}^{p'}\,G{}\!\,_{;}{}_{q}{}^{q} +
    G\,G{}\!\,_{;}{}_{p}{}^{p}{}_{q'}{}^{q'} \right)  \cr
&& - 4\,\left( G{}\!\,_{;}{}_{p}\,G{}\!\,_{;}{}_{q'}{}^{p}{}^{q'} +
    G{}\!\,_{;}{}^{p'}\,G{}\!\,_{;}{}_{q}{}^{q}{}_{p'} \right)  \cr
&& + R\,G{}\!\,_{;}{}_{p'}\,G{}\!\,_{;}{}^{p'} +
  {R'}\,G{}\!\,_{;}{}_{p}\,G{}\!\,^{;}{}^{p} \cr
&& - 2\,\left( R\,G{}\!\,_{;}{}_{p'}{}^{p'} +
{R'}\,G{}\!\,_{;}{}_{p}{}^{p} \right)
     G
+  {\frac{1}{2}} R\,{R'} {G^2}  \;.
\end{eqnarray}
\label{generalnoise}
\ees
Note that the superscript $+$ on $G^+$ has been omitted in the above equations for notational simplicity.  Primes on indices denote tensor indices at the point $x'$ and unprimed ones denote indices at the point $x$.  $R_{ab}$ is the Ricci tensor evaluated at the point $x$, $R_{c'\,d'}$ is the Ricci tensor evaluated at the point $x'$, $R$ is the scalar curvature evaluated at the point $x$, and $R'$ is the scalar curvature evaluated at the point $x'$.

The definition~\eqref{NKdef} of the noise kernel immediately implies that it is symmetric under interchange of the two spacetime points and the corresponding pairs of indices so that
 \be N_{abc'd'}(x,x') = N_{c'd'ab}(x',x) \;. \ee
The noise kernel obeys other important properties as well.  These have been proven in Refs.~\cite{martin99a,martin99b}, so we just state them here.
The first property, which is clear from~\eqref{NKdef}, is that the following conservation laws must hold:
\be  \nabla^a N_{abc'd'} = \nabla^b N_{abc'd'}= \nabla^{c'} N_{abc'd'} = \nabla^{d'} N_{abc'd'} = 0 \;.  \label{conservation} \ee
The second property which must be satisfied because the field is conformally invariant is that the partial traces must vanish, that is
\be {N^a}_{ a c'd'} = N^{\ \  c'}_{a b\ c'} = 0 \;. \label{partial-traces} \ee
A third important property is that the noise kernel is positive semidefinite, namely
\begin{equation}
\int d^4 x \sqrt{-g(x)} \int d^4 x' \sqrt{-g(x')} \, f^{ab}(x) N_{abc'd'}(x,x') f^{c'd'}\!(x') \geq 0 ,
\end{equation}
for any real tensor field $f^{ab}(x)$.

Finally the noise kernel for the conformally invariant field
has a simple scaling behavior under conformal transformations.
In the Appendix
two proofs are given which show that under a conformal transformation between two conformally related $D$-dimensional spacetimes with metrics of the form
$\tilde{g}_{ab} = \Omega^2(x)\, g_{ab}$ and conformally related states, the noise
kernel transforms as
\begin{equation}
\tilde{N}_{abc'd'}(x,x') =  \Omega^{2-D}(x)\, \Omega^{2-D}(x')\, N_{abc'd'}(x,x')  \;.
\label{eq:rescaling1}
\end{equation}

\section{Gaussian approximation in the Optical Schwarzschild spacetime}
\label{sec:gaussian}

As discussed in the introduction, we want to compute the noise kernel in a background Schwarzschild spacetime for the conformally invariant scalar field when
the points are separated.  For an arbitrary separation it would be necessary to do this numerically.  However, if the separation is small then it is possible
to use approximation methods to compute the Wightman function analytically and from that the noise kernel. For a conformally invariant field a significant
simplification is possible because the Green function and the resulting noise kernel can be computed in the optical Schwarzschild spacetime (which is conformal to
Schwarzschild spacetime) and then conformally transformed to Schwarzschild spacetime.  A similar calculation was done by Page~\cite{page} for the stress tensor expectation value of a conformally invariant scalar field.  He first calculated the Euclidean Green function for the field in a thermal state using
a Gaussian approximation.  Then the stress tensor was computed and conformally transformed to Schwarzschild spacetime.  We shall use Page's approximation for
the Euclidean Green function to obtain an approximation for the Wightman Green function and then compute the noise kernel using that approximation.

\subsection{Gaussian approximation for the Wightman Green function}
\label{sec:wightman}

In order to use Page's approximation we must relate the Euclidean Green function in a static spacetime to the Wightman function.
To do so we begin by noting that the Wightman function can be written in terms of two other Green functions~\cite{bir-dav-book}, the Hadamard function $G^{(1)}(x,x')$ and the Pauli-Jordan function $G(x,x')$:
\bes
\bea  G^{+}(x,x') &=& \frac{1}{2} \left[ G^{(1)}(x,x') + i G(x,x') \right]  \label{Wightman-def-2} \\
 G^{(1)}(x,x') &\equiv& \langle \{\phi(x), \phi(x')\} \rangle  \label{Hadamard-def} \\
 i\, G(x,x') &\equiv& \langle [\phi(x), \phi(x')] \rangle \;.
\eea
\ees
\noindent As discussed in the Introduction, we restrict our attention in this paper to spacelike and timelike separations of the points.
In general $G(x,x') = 0$ for spacelike separations. Furthermore, in the optical Schwarzschild
spacetime, $G(x,x') = O[(x-x')^4]$ for timelike separations of the points.  To see this consider the general form of the Hadamard
expansion for $G(x,x')$ which is~\cite{aeh}
\be G(x,x') =  -\frac{u(x,x')}{4 \pi} \delta(-\sigma)  + \frac{v(x,x')}{8 \pi}  \theta(-\sigma) \;, \label{G-expansion} \ee
with $\sigma(x,x')$ defined to be one-half the square of the proper distance along the shortest geodesic connecting the two points.
In~\cite{AndHu04} it was shown that in Schwarzschild spacetime $v(x,x') = O[(x-x')^4]$.  Since the Green function in the optical spacetime
can be obtained from that in Schwarzschild spacetime by a simple conformal transformation, the same must be true of the quantity $v(x,x')$.
Thus, so long as we work only to $O[(x-x')^2]$ and restrict our attention to points which are either spacelike or timelike separated, then in the optical Schwarzschild spacetime
\be G^+(x,x') = \frac{1}{2} G^{(1)}(x,x')  + O[(x-x')^4] \;.  \label{Wightman-Hadamard}  \ee

The Hadamard Green function can be computed using the Euclidean Green function in the following way.  First, define the Euclidean time as
\be \tau \equiv i t  \;. \label{Euclidean-time} \ee
Then the metric in a static spacetime takes the form
\be ds^2 = g_{\tau \tau}(\vec{x}) d \tau^2  + g_{i j}(\vec{x}) dx^i dx^j \;. \ee
The Euclidean Green function obeys the equation
\be (\Box  - \frac{1}{6}R) G_E(x,x') = -\frac{\delta(x-x')}{\sqrt{g(x)}}  \;.  \ee
Because the spacetime is static $G_E$ will be a function of $(\Delta \tau)^2 = (\tau - \tau')^2$.  It is possible to obtain the Feynman Green function
$G_F(x,x')$ by making the following transformation~\cite{gac}:
\be (\Delta \tau)^2 \rightarrow - (\Delta t)^2 + i \epsilon  \;,  \label{continuation} \ee
under which
\be G_E(x,x') \rightarrow i G_F(x,x') \;. \label{GE-GF} \ee
Writing the Feynman Green function in terms of the Hadamard and Jordan functions \cite{bir-dav-book},
\be G_F(x,x') = - \frac{1}{2} i \, G^{(1)}(x,x') + \frac{1}{2} \left[ \theta(t-t') - \theta(t'-t) \right] G(x,x') \;. \ee
one finds that Eq.~\eqref{Wightman-Hadamard} becomes
\be G^+(x,x') = -{\rm Im} \, G_F(x,x')  + O[(x-x')^4] \;.  \label{Wightman-Feynman}  \ee

As mentioned above, Page made use of the DeWitt-Schwinger expansion to obtain his approximation for the Euclidean Green function.  Before displaying
his approximation it is useful to discuss two quantities which appear in that expansion.  For a more complete discussion see Ref.~\cite{christensen}.
The fundamental quantity out of which everything is built is Synge's world function $\sigma(x,x')$, which is defined to be one-half the square of the
proper distance between the two points $x$ and $x'$ along the shortest geodesic connecting them.  It satisfies the relationship
\be \sigma(x,x') = \frac{1}{2} g_{ab}(x) \, \sigma^{; a}(x,x')\, \sigma^{; b}(x,x') \;,\label{sigma-def} \ee
and it is traditional to use the notation
\be \sigma^a \equiv \sigma^{;a} \;. \ee
As shown in Ref.~\cite{christensen}, it is possible to expand biscalars, bivectors, and bitensors in powers of $\sigma^a$ in an arbitrary spacetime
about a given point $x$.  Then the coefficients in that expansion are evaluated at the point $x$.
For example
\begin{eqnarray}
\sigma_{;ab}(x,x') &=& g_{ab}(x) -\frac{1}{3} R_{acbd}(x) \,
\sigma^c(x,x') \sigma^d(x,x') + \cdots  \;.  \label{sigmaab}
\end{eqnarray}
Examination of this expansion shows that to zeroth order in $\sigma^a$
\be \sigma_{;abc} = 0  \;.  \label{sigmaabc} \ee

The second quantity we shall need is
\bes
\label{u-Delta-def}
\bea U(x,x') &\equiv& \Delta^{1/2}(x,x')  \,, \\
  \Delta(x,x') &\equiv& -\frac{1}{\sqrt{-g(x)} \, \sqrt{-g(x')}} \,  {\rm det}\left(- \sigma_{;\,a \, b\,'} \right)\,. \label{eq:Delta}
\eea
\ees
Note that covariant derivatives at the point $x'$ commute with covariant derivatives at the point $x$.
Two important properties of $U(x,x')$ are
\bes
\bea  U(x,x) &=& 1    \label{Ucondition} \\
   \left( \ln U \right)_{; a} \sigma^a &=& 2 - \frac{1}{2} \Box \, \sigma  \;.  \label{Ueq}
\eea
\ees
One can also expand $U$ in powers of $\sigma^a$
with the result that~\cite{christensen}
\bea U(x,x') &=& 1
 + \frac{1}{12} R{}_{a}{}_{b} \,\sigma^a \sigma^b
  - \frac{1}{24} R{}_{a}{}_{b}{}_{;}{}_{c} \, \sigma^a \sigma^b \sigma^c , \nonumber \\
& +& \frac{1}{1440}\left(
18\,R{}_{a}{}_{b}{}_{;}{}_{c}{}_{d} +
5\,R{}_{a}{}_{b}\,R{}_{c}{}_{d} +
  4\,R{}_{p}{}_{a}{}_{q}{}_{b}\,R{}_{c}{}^{q}{}_{d}{}^{p}\right) \, \sigma^a \sigma^b \sigma^c \sigma^d + O[(\sigma^a)^5] .
 \label{Apx-VanDseries}
\eea

The above definitions, properties and expansions apply to arbitrary spacetimes.
Given any static metric, one can always transform it to an ultra-static one, called the \emph{optical metric}, by a conformal transformation. This kind of metric is of the form
\begin{eqnarray}
ds^2 = -dt^2 + g_{ij}(\vec{x}) \, d x^i d x^j  \,, \label{metric1}
\end{eqnarray}
with the metric functions $g_{ij}$ independent of the time $t$.  In this case
Synge's world function is given by
\begin{eqnarray}
\sigma(x,x') = \frac{1}{2}\left(-(t-t')^2 + \mathbf{r}^2\right)  \,,\label{sigma-ultrastatic}
\end{eqnarray}
with
\begin{eqnarray}
\mathbf{r} \equiv \sqrt{2\; ^{(3)}\sigma} \;. \label{rdef}
\end{eqnarray}
The quantity
$^{(3)}\sigma$ is the part of $\sigma$ that depends
on the spatial coordinates.
Note that we use $\mathbf{r}$ (with bold roman font) to denote the quantity
in Eq.\eqref{rdef} while $r$
(with normal italic font) denotes the radial coordinate.
Some useful properties of $\mathbf{r}$ are
\bes
\begin{eqnarray}
\label{a2}\nabla_i \mathbf{r} &=& \frac{\;^3\sigma_i}{\mathbf{r}} \,, \\
\label{a3}\nabla^2 \mathbf{r} &=& \frac{\;^3\sigma^i_i - 1}{\mathbf{r}} \,, \\
\label{a4}\nabla_i \mathbf{r} \; \nabla^i
\mathbf{r} &=& \frac{\;^3\sigma_i
\;^3\sigma^i}{2\;^3\sigma} = 1 \,, \\
\label{a5}2 \left({}^{(3)}\!\Delta^{1/2}\right)_{,i}\;\; \nabla^i
\mathbf{r} &=& \left(\frac{3}{\mathbf{r}} -
\nabla^2 \mathbf{r}\right) {}^{(3)}\!\Delta^{1/2} \,,
\end{eqnarray}
\ees
where $\nabla^2 = \nabla^i\nabla_i $.
Note that from Eqs.~(\ref{eq:Delta}) and (\ref{sigma-ultrastatic}) one can easily see that for an ultra-static spacetime the Van-Vleck determinant $^{(3)} \Delta$ for the spatial metric $g_{ij}$ coincides with the Van-Vleck determinant $\Delta$ for the full spacetime. (The advantage of using $^{(3)} \! \Delta$ rather than $\Delta$ is that, although noncovariant, it is valid for arbitrary time separations, and one only needs to expand in powers of $\mathbf{r}$.)

The optical Schwarzschild metric
\begin{eqnarray}
ds^2 =   -dt^2 + \frac{1}{\left( 1 - \frac{2 M}{r} \right)^2} d r^2
+ \frac{r^2}{1 - \frac{2M}{r}} \, \left( d \theta^2 + \sin^2 \theta d \phi^2 \right) \;, \label{optical-Sch-metric}
\end{eqnarray}
is of the form~\eqref{metric1} and is conformally related to the Schwarzschild metric in standard coordinates with a conformal factor
\be \Omega^2 = \left( 1 - \frac{2 M}{r} \right) \;.  \label{Omega-def} \ee
For this metric Page~\cite{page} used a Gaussian approximation to obtain an expression for the Euclidean Green function in a thermal state at
the temperature
\be T =  \frac{\kappa}{2 \pi} \;. \label{eq:temperature}
\ee
This expression is valid for any temperature; however, if
\be \kappa = \frac{1}{4 M} \;, \label{eq:surf_grav} \ee
then the temperature is that of the black hole in the Schwarzschild spacetime which is conformal to the optical metric~\eqref{optical-Sch-metric}.
In this case the state of the field corresponds to the Hartle-Hawking state, which is regular on the horizon.  Page found that~\cite{page}
\begin{eqnarray}
G_{E}(\Delta \tau,\vec{x},\vec{x}\,') = \frac{\kappa \,\sinh(\kappa
\mathbf{r})}{8\pi^2 \mathbf{r}\, [\cosh(\kappa \mathbf{r}) - \cos (\kappa
\Delta \tau)]}  \, U(\Delta \tau,\vec{x},\vec{x}\,') \;. \label{GE-gauss-1}
\end{eqnarray}
Analytically continuing to the Lorentzian sector using the prescriptions~\eqref{continuation} and~\eqref{GE-GF}, and
substituting the result into Eq.~\eqref{Wightman-Feynman} gives
\begin{eqnarray}
G^+(\Delta t,\vec{x},\vec{x}\,') = \frac{\kappa \sinh\kappa
\mathbf{r}}{8 \pi^2 \mathbf{r}[\cosh(\kappa \mathbf{r}) - \cosh( \kappa
\Delta t)])}  \, U(\Delta t,\vec{x},\vec{x}\,')\;. \label{Gwgauss}
\end{eqnarray}

To determine the accuracy of this approximation we can
substitute the above expression
into the equation satisfied by $G^+$ which is
\be \Box \, G^+(x,x') - \frac{R}{6} G^+(x,x') = 0 \;. \label{GWightman-eq} \ee
The accuracy of the Gaussian approximation in the optical Schwarzschild metric will
be determined by the lowest order in $(x-x')$ at which Eq.~\eqref{GWightman-eq} is not satisfied.
Applying the differential operator for the metric~\eqref{optical-Sch-metric}  and using Eqs.~\eqref{a4}-\eqref{a5}, one finds after some calculation that
\begin{eqnarray}
\label{eq:conffield}\left(\Box - \frac{1}{6}R\right)G^+(x,x') =
 \frac{\kappa\,\sinh(\kappa\,\mathbf{r})}
   {\mathbf{r} \left[\cosh (\kappa\,\mathbf{r})-\cosh (\kappa\,\Delta t) \right]}\,\left(\Box- \frac{1}{6} R\right) \, U(x,x')
\end{eqnarray}
If Eq.~\eqref{Apx-VanDseries} is substituted into Eq.~\eqref{eq:conffield} and Eqs.~\eqref{sigmaab}-\eqref{sigmaabc} are used, then one finds
\begin{eqnarray}
\label{eq:deltainfe} \left(\Box -
\frac{R}{6}\right)\, U(x,x') &=& Q_0 + Q_{p}\sigma^p +
Q_{pq}\sigma^p \sigma^q + \cdots
\end{eqnarray}
with
\bes
\bea
Q_0 &=& 0  \,, \\
Q_a \sigma^a &=& \sigma^a \, {{G_a}^b}_{;b} = 0  \,, \\
  Q_{ab} \, \sigma^a \, \sigma^b &=&
\frac{1}{360} \left( 9R_{;ab} + 9{R_{ab;c}}^{c} - 24{{R_{ac;}}^{c}}_{b} - 12R_{ac}{R_b}^c \right. \nonumber \\
& & \left. + 6R^{cd}R_{cadb} + 4R_{acde}{R_b}^{edc} + 4R_{acde}{R_b}^{cde} \right) \sigma^a \sigma^b \,. \label{eq:secondorderrhsdelta}
\eea
\ees
Here $G_{ab}$ is the Einstein tensor.
For the optical Schwarzschild metric~\eqref{optical-Sch-metric}, $Q_{ab}\, \sigma^a \sigma^b = 0$.  Thus, Eq.~\eqref{eq:deltainfe} is zero to $O[(x-x')^2]$.  Since $\Box$ is a second
order derivative operator, this means that the Gaussian approximation for $G^+(x,x')$, whose leading order behavior is $G^+(x,x') \sim (x-x')^{-2}$, is accurate through $O[(x-x')^2]$.
Note that this approximation is valid for arbitrary temperature since Eq.~\eqref{eq:conffield} holds for arbitrary values of $\kappa$.

It is important to emphasize that the order of accuracy obtained here is for the Schwarzschild optical metric~\eqref{optical-Sch-metric}.  Because the Gaussian approximation
is equivalent to the lowest order term in the DeWitt-Schwinger expansion, it is only guaranteed to be accurate to leading order in $x-x'$ in a general spacetime.

\subsection{Order of validity of the noise kernel}
\label{sec:validity}

In Sec.~\ref{sec:general} an expression for the noise kernel is given in terms of covariant derivatives of the Wightman function.  In each term there is a product of Wightman
functions and varying numbers of covariant derivatives.  The accuracy of the Gaussian approximation for the Wightman function can be used to estimate the order of accuracy of the noise kernel.  First, recall that the leading order of the Wightman function goes like $(x-x')^{-2}$.  Since there is a maximum of four derivatives acting on a product of Wightman functions, one expects that at leading order the noise kernel will go like $(x-x')^{-8}$.  Since the Gaussian approximation to the Wightman function
in the optical Schwarzschild spacetime is accurate through terms of order $(x-x')^2$, it is clear from Eq.~(\ref{generalnoise}) that our expression for the noise kernel should be accurate up to and including terms of order $(x-x')^{-4}$.

\section{Computation of the Noise Kernel}
\label{sec:computation-nk}

In this section we discuss the computation of the noise kernel in two different but related cases.  In both the field is in a thermal state at an arbitrary temperature $T$ and the points are separated in a non-null direction.  The first case considered is flat space where the calculation of the noise kernel is exact.  In the second case an approximation to the noise kernel is computed for the optical Schwarzschild metric~\eqref{optical-Sch-metric}.  The result is then conformally transformed to Schwarzschild spacetime using Eq.~\eqref{eq:rescaling1}.

\subsection{Hot Flat Space}
\label{sec:hfs}

In flat space the function $U(x,x')$ is exactly equal to one.  Examination of Eq.~\eqref{eq:conffield} then shows that the expression for $G^+(x,x')$
in Eq.~\eqref{Gwgauss} is exact so long as the points are separated in a non-null direction.  This expression can be substituted into Eqs.~\eqref{general-noise-kernel}
and~\eqref{generalnoise} to obtain an exact expression for the noise kernel.  Here the quantity $\mathbf{r}$ takes on the following simple form in Cartesian coordinates and components:
\be \mathbf{r} = \sqrt{(x-x')^2 + (y-y')^2 + (z-z')^2} \;. \label{r-flat-space}  \ee
The only subtlety which one must be aware of is that the point separation must be arbitrary before the derivatives are computed.  Once they are computed, then any point separation that one desires can be used.

All components of the noise kernel have been calculated when the points are separated in a non-null direction.  Both conservation and the vanishing of the partial traces have
been checked.  Due to the length of many of the components, we just display one of them here:
\bea
  N_{ttt't'} &=& \frac{\kappa ^2 \sinh ^2(\kappa \mathbf{r} )}{192 \pi ^4 \mathbf{r}^6 (\cosh (\kappa \Delta t )-\cosh (\kappa \mathbf{r} ))^2} \nonumber \\
                 & & \nonumber \\
                 & &  + \frac{\kappa ^3 \sinh (\kappa \mathbf{r} )}{96 \pi^4 \mathbf{r}^5 (\cosh (\kappa \Delta t )-\cosh (\kappa \mathbf{r} ))^3}
                        \left[ 1 - \cosh (\kappa \Delta t ) \cosh (\kappa \mathbf{r} ) \right] \nonumber \\
                 & & \nonumber \\
                 & & + \frac{\kappa ^4} {192 \pi ^4 r^4 (\cosh (\kappa \Delta t )-\cosh (\kappa \mathbf{r} ))^4}
                         \left[ 2 - 2 \cosh (\kappa \Delta t ) \cosh (\kappa \mathbf{r} ) \right. \nonumber \\
                 & & \left. - \cosh^2 (\kappa \mathbf{r} ) + \cosh^2 (\kappa \Delta t ) \cosh (2 \kappa \mathbf{r} ) \right] \nonumber \\
                 & & \nonumber \\
                 & &  + \frac{\kappa ^5  \sinh (\kappa \mathbf{r} ) }{288 \pi ^4 \mathbf{r}^3 (\cosh (\kappa \Delta t )-\cosh (\kappa \mathbf{r} ))^4}
                         \left[ 2 \cosh (\kappa \Delta t) -  \cosh (2 \kappa \Delta t) \cosh (\kappa \mathbf{r} ) \right. \nonumber \\
                 & &  \left. - \cosh (\kappa \Delta t) \cosh^2 (\kappa \mathbf{r} ) \right] \nonumber \\
                 & & \nonumber \\
                 & & -\frac{\kappa^6}{576 \pi ^4 \mathbf{r}^2 (\cosh (\kappa \Delta t )-\cosh (\kappa \mathbf{r} ))^6}
                       \left[12 - 6 \cosh^2 (\kappa \Delta t) \right. \nonumber \\
                 & & \left. + \cosh^4 (\kappa \Delta t) - 12 \cosh (\kappa \Delta t ) \cosh (\kappa \mathbf{r} ) + \cosh^3 (\kappa \Delta t ) \cosh (\kappa \mathbf{r} ) \right. \nonumber \\
                 & & \left. -18 \cosh^2 (\kappa \mathbf{r} ) + 12 \cosh^2 (\kappa \Delta t ) \cosh^2 - 2 \cosh^4 (\kappa \Delta t ) \cosh^2 (\kappa \mathbf{r} )  (\kappa \mathbf{r} ) \right. \nonumber \\
                 & & \left. + 17 \cosh (\kappa \Delta t) \cosh^3 (\kappa \mathbf{r} ) + 3 \cosh^3 (\kappa \Delta t) \cosh^3 (\kappa \mathbf{r} ) - \cosh^4 (\kappa \mathbf{r} )\right. \nonumber \\
                 & & \left. - 6 \cosh (2 \kappa \Delta t) \cosh^4 (\kappa \mathbf{r} ) - \cosh (\kappa \Delta t) \cosh^5 (\kappa \mathbf{r} ) \right] \;. \label{NKhfs}
\eea

\subsection{Schwarzschild Spacetime}
\label{sec:computation-nk-Sch}

The calculation of the noise kernel in the optical Schwarzschild
metric~\eqref{optical-Sch-metric} proceeds in the same way as the flat space
calculation in Sec.~\ref{sec:hfs}.
That is, one simply substitutes the expression~\eqref{Gwgauss} for $G^+(x,x')$, which is now approximate rather than exact, into Eqs.~\eqref{general-noise-kernel}-\eqref{generalnoise} and computes the various derivatives and curvature tensors, again with an arbitrary separation of the points.  After the derivatives are computed, the specific separation of the points which is of interest may be taken. Because the expression for $G^+(x,x')$ is approximate, one must expand the result in powers of $(x-x')$ and, as discussed at the end of Sec.~\ref{sec:gaussian}, the result should be truncated at order $(x-x')^{-4}$.

The expansion in powers of $(x-x')$ can be consistently done for all contributions to the noise kernel and the results of such an expansion are shown below in Sec.~\ref{sec:low_temp}.
However, if the separation of the points is only in the time direction, then it is possible to obtain a result valid for arbitrary values of $\kappa$ (and, hence, of $\kappa \Delta t$). This can be achieved by treating exactly the prefactor in Eq.~\eqref{Gwgauss}, which contains all the dependence on $\kappa$, while expanding
the quantity $U(x,x')$ and its derivatives in powers of $(x-x')$.
There are two reasons why this works.  The first is that, as can be seen from Eq.~\eqref{eq:conffield}, what is keeping $G^+(x,x')$ in Eq.~\eqref{Gwgauss} from being exact is the fact that for the optical Schwarzschild metric
$\Box_x U(x,x')$ is not exactly zero but only vanishes to order $(x-x')^2$. So in some sense the function that multiplies $U$ in Eq.~\eqref{Gwgauss} can be treated as exact. Secondly, although exact analytic expressions for the function $\mathbf{r}$ and its derivatives in terms of simple functions are not known for an arbitrary splitting of the points, in the limit that the points are separated only in the time direction such expressions are known. Therefore, it should be possible to treat these terms exactly when the final point separation is in the time direction.  It is worth noting that it is consistent to have a quasi-local expansion in which $\Delta t$ is in an appropriate sense small (as specified in footnote~\ref{foot:expansion}) and yet to have $\kappa |\Delta t| \gtrsim 1$. The reason is that the scale over which the geometry varies significantly in the optical metric is $O(M)$.  For the Hartle-Hawking state $\kappa = 1/(4 M)$ and the validity of the quasi-local expansion would start to break down for a $\Delta t$ such that $\kappa |\Delta t| \sim 1$, but the Hartle-Hawking state is a very low temperature state for a macroscopic black hole.  Therefore, one can have temperatures which are well below the Planck temperature and still have $\kappa \Delta t \gg 1$.
Furthermore, even when $\kappa |\Delta t| \ll 1$ it is sometimes useful to have the exact dependence on $\kappa \Delta t$. An example illustrating this point is the Rindler limit of Schwarzschild spacetime with the field in the Hartle-Hawking state, which is discussed in Sec.~\ref{sec:arb_temp}.

To compute the noise kernel using the above method it is necessary to find expansions for both $\mathbf{r}$ and $U(x,x')$ in powers of $(x-x')$.  For the former it is easier to work with
the quantity $\sigma(x,x')$ which is one-half the square of the proper distance between $x$ and $x'$ along the shortest geodesic that connects them.  The relationship
between $\sigma$ and $\mathbf{r}$ is given by Eq.~\eqref{sigma-ultrastatic}.   Furthermore, since the metric~\eqref{optical-Sch-metric} is also spherically
symmetric, $\sigma$ can only depend on the angular quantity
\begin{equation} \cos(\gamma) \equiv \cos \theta  \, \cos \theta' + \sin \theta \, \sin \theta' \, \cos(\phi-\phi')  \label{cosg} \,, \end{equation}
where $\gamma$ is the angle between $\vec{x}$ and $\vec{x}{\,'}$.  It turns out to be convenient to write $\sigma$ in terms of the quantity
\be \eta \equiv \cos \gamma - 1 \;. \ee
Then for points that are sufficiently close together one can use the expansion
\be
\sigma(x,x') = - \frac{1}{2} (t-t')^{2}  + \sum_{j,k} w_{jk}(r) \, \eta^j \, (r-r')^k
\;, \label{sigma-sum}
\ee
with the sums over $j$ and $k$ starting at $j = 0$ and $k = 0$ respectively.
For the metric~\eqref{optical-Sch-metric}, Eq.~\eqref{sigma-def} has the explicit form
\be \sigma = \frac{1}{2} \left[ - \left(\frac{\partial \sigma}{\partial t} \right)^2 + \left(1-\frac{2M}{r} \right)^2 \left(\frac{\partial \sigma}{\partial r} \right)^2
  - \frac{1}{r^2}\,\left(1-\frac{2M}{r} \right)  \left(\frac{\partial \sigma}{\partial \eta} \right)^2 (2 \eta + \eta^2)  \right]  \;.  \label{sigma-eq-2}
\ee
Substituting the expansion~\eqref{sigma-sum} into Eq.~\eqref{sigma-eq-2} and equating powers of $(x^a - x^{a'})$, one finds that
\bea  \sigma(x,x')  &=& -\frac{(\Delta t)^2}{2} + \frac{(\Delta r)^2}{2 f^2} - \frac{r^2 \eta }{f} + (\Delta r)^3 \left( \frac{1}{2 r f^3} -\frac{1}{2 r f^2} \right) \nonumber \\
& & + \eta \, \Delta r \left( \frac{3 r}{2 f} - \frac{r}{2 f^2} \right) + O[(x-x')^4] \;,\label{sigma-solution}  \eea
with $\Delta r \equiv r - r'$ and
\be f \equiv 1 - \frac{2 M}{r}  \;. \label{f-def} \ee

Because of the form of Eq.~\eqref{Ueq} an expansion for $U(x,x')$ can be found by writing
\be \ln U(x,x') =  \sum_{j,k} u_{jk}(r) \, \eta^j \, (r-r')^k \;. \label{U-sum}\ee
It can be seen from Eqs.~\eqref{sigma-ultrastatic} and~\eqref{u-Delta-def} that
for a spacetime with metric~\eqref{optical-Sch-metric} $U$ is time independent.
If Eq.~\eqref{U-sum} is substituted into Eq.~\eqref{Ueq},  Eqs.~\eqref{Ucondition} and~\eqref{sigma-solution} are used, and the result is expanded
in powers of $(x-x')$, then one finds that
\be U(x,x') = 1 + \frac{(\Delta r)^2}{8 r^2} \left(1 - \frac{2}{3 f} - \frac{1}{3 f^2} \right)
+ \eta \left(\frac{f}{4}  - \frac{1}{3} + \frac{1}{12 f} \right) + O[(x-x')^3] \;. \label{U-solution}
\ee
Since the leading order of the prefactor multiplying $U$ in Eq.~\eqref{Gwgauss} is $(x-x')^{-2}$, we need to calculate $U(x,x')$ through $O[(x-x')^4]$. That way we can obtain the Wightman function through $O[(x-x')^2]$, which is consistent
with the order to which the Gaussian approximation was shown to be valid in Sec.~\ref{sec:gaussian}.
One can compute $U(x,x')$ to the required order by substituting the expansions~\eqref{sigma-solution} for $\sigma$ and~\eqref{U-solution} for $U$ into Eq.~\eqref{Ueq}.
To obtain a final expression for $U$ valid through $O[(x-x')^4]$ it is necessary to
have the expansion for $\sigma$ containing terms through  $O[(x-x')^6]$.
[As an illustration, we have shown the results through quadratic order in Eqs.~\eqref{sigma-solution} and \eqref{U-solution}.]

Using Eq.~\eqref{sigma-ultrastatic} an expansion for the quantity $\mathbf{r}$ can be obtained from the expansion for $\sigma(x,x')$.  This along with the expansions for $U(x,x')$ can be substituted into Eq.~\eqref{Gwgauss} to obtain an expansion for the Wightman function $G^+(x,x')$.  That in turn
can then be substituted into the expressions~\eqref{generalnoise} for the noise kernel and the derivatives can be computed.  As discussed in Sec.~\ref{sec:validity}, one should keep terms through $O[(x-x')^{-4}]$ since this is the highest order for which the Gaussian approximation for the noise kernel is valid.  To obtain the noise kernel for
Schwarzschild spacetime one then uses the conformal transformation~\eqref{eq:rescaling1} with $\Omega^2(x) = 1-2M/r$.  Finally, the coordinate $r'$ is written as $r' = r - (r-r')$ in order to expand the resulting expression in powers of $(r-r')$ through quartic order.

\subsubsection{Arbitrary Temperature}
\label{sec:arb_temp}

Following the method described above and using the exact expression for the $\kappa$-dependent prefactor in Eq.~\eqref{Gwgauss}, we have computed several components of the noise kernel for points split in the time direction when $\kappa \Delta t$ is not assumed to be small.  The result for the $N_{t~t'~}^{~t~t'}$ component is
\bea N_{t~t'~}^{~t~t'} &=& \frac{1}{1728 \pi ^4 f^4 } \Bigg[ \kappa^8
   \frac{ \left(2 \cosh ^2(\kappa \Delta t)-\cosh (\kappa \Delta t )+26\right)}{(\cosh (\kappa \Delta t )-1)^4} \nonumber \\
                                         & & \nonumber \\
                                         & & +\frac{\kappa^6 }{4 r^2}
   \frac{ (1-f)^2 (1-4 \cosh (\kappa \Delta t))}{(\cosh (\kappa \Delta t )-1)^3} \nonumber \\
                                         & & \nonumber \\
                                         & &  +\frac{\kappa^4}{8 r^4}
   \frac{(1-f)^2 (1-2f+3f^2) }{(\cosh (\kappa \Delta t )-1)^2} \Bigg]  \;. \label{kappa-r-large}
\eea

There are two ways to take the flat space limit of this result.
One possibility is to take $f \rightarrow 1$ in Eq.~\eqref{kappa-r-large}.
This limit reduces to the corresponding expression for the exact noise kernel in flat space at arbitrary temperature, whose spatial coincidence limit can be obtained from Eq.~\eqref{NKhfs} by taking the limit $\mathbf{r} \rightarrow 0$.

A second possibility for obtaining the flat space limit is to realize that the
geometry of the near-horizon region is that of Rindler spacetime and in the
limit of large Schwarzschild radius this holds for an arbitrarily large region.
Indeed, if one introduces the new coordinates
\begin{align}
\begin{split}
\label{eq:ibx}
\xi & = 4M \sqrt{r/2M - 1}\, , \\
T & = \frac{t}{4M} \, ,
\end{split}
\end{align}
in the near-horizon region, characterized by $|r/2M - 1| \ll 1$, the standard
Schwarzschild metric reduces to
\be
ds^2 \approx -\xi^2 dT^2 + d\xi^2 + dx_\perp^2 \, ,
\ee
where $dx_\perp^2 = 4M^2 (d\theta^2 + \sin^2\theta\, d\phi^2)$ becomes the metric of a Euclidean plane (say, tangent to $\theta=\phi=0$) when $M \to \infty$.
In the new coordinates, the near-horizon condition corresponds to $\xi \ll 4M$. Therefore, in the limit $M \to \infty$ one recovers the full Minkowski spacetime in Rindler coordinates. Rewriting Eq.~\eqref{kappa-r-large} in terms of $\xi$ and $T$, and taking the limit $M \to \infty$, we get
\be
N_{T~T'~}^{~T~T'}
= \frac{1}{4\pi^4} \frac{1}{\big(2 \, \xi \, \sinh(\Delta T/2) \big)^8} \, . \label{N-Rindler}
\ee
This result agrees with the flat space calculation of the noise kernel in
Ref.~\cite{martin00} if one takes into account that our definition of the noise
kernel is four times their definition. In order to do the comparison one needs
to consider Eqs.~(3.10), (4.9) and (4.12) in Ref.~\cite{martin00} and rewrite
their results in terms of Rindler coordinates by using their relation to
inertial coordinates ($x^0 = \xi \sinh T,\ x^1 = \xi \cosh T$); in particular,
this implies that for pairs of points with equal $\xi$ the Minkowski interval is
given by $(x-x')^2 = 4 \, \xi^2 \sinh^2 (\Delta T/2)$.
It also requires transforming the tensor components accordingly using the Jacobian of the coordinate transformation.

It should be noted that in order to get the Minkowski vacuum in this limit, one needs to consider the Hartle-Hawking state, which is regular on the horizon, for the black hole. In that case $\kappa$ is tied to $M$ as given by Eq.~\eqref{eq:surf_grav}, so that $\kappa\Delta t = \Delta T$.
It is, therefore, important to have an expression valid for arbitrarily large $\kappa\Delta t$, because this guarantees that the exact Rindler result is obtained, rather than an approximate expansion valid only up to some order for small $\Delta T$.
(The condition for the validity of the quasi-local expansion is
$\sigma/R_S^{\,2} \approx 8 \, (\xi / 4M)^2 \sinh^2 (\Delta T/2) \ll 1$, which is fulfilled for any values of $\xi$ and $\Delta T$ as $M \to \infty$.)

The two different ways of obtaining the flat space limit described above provide
(partially independent) nontrivial checks of our result.  We have also checked one of the partial traces which involve the $N_{t~t'~}^{~t~t'}$ component and have shown that it vanishes to the appropriate order.  Finally, we have partially checked one of the conservation conditions by initially considering an arbitrary separation of the points, computing the relevant derivatives, and then taking the limit that the separation is only in the time direction.  It was only shown that the conservation condition is satisfied to the second highest order used in the computation.  In terms of the expansion of $U(x,x')$ discussed above this corresponds to order $(x-x')^2$.  We are confident that the conservation condition is also satisfied to the highest order used in the computation, $O[(x-x')^4]$, but due to the large number of terms involved, we have not shown this explicitly.

\subsubsection{Moderate and low temperature}
\label{sec:low_temp}

If one is interested in point separations such that $\kappa \Delta t \ll1$ and $\kappa \mathbf{r} \ll1$, then it is useful to expand the Wightman function in powers of $(x-x')$ before substituting it into the expressions~\eqref{General-Noise}.  The general expression~\eqref{GE-gauss-1} for $G^+(x,x')$ in the Gaussian approximation can be expanded as~\cite{PH03}
\begin{eqnarray}
G^+(x,x') &=&  \frac{1}{8\,{{\pi
}^2}}  \left[ \frac{1}{\sigma} +  {\frac{{{\kappa}^2}}{6}}
  - {\frac{{{\kappa}^4}}{180}} \left( 2\,{{{(\Delta t)}}^2} + \sigma \right)
+O\left[(x-x')^4\right] \right] \, U(x,x') \label{Opt-Wightman} \, .
\end{eqnarray}
Since $U(x,x') = 1 + O[(x-x')^2]$, the terms within the square brackets in Eq.~\eqref{Opt-Wightman} have been kept through $O[(x-x')^2]$, which is consistent with the order to which the Gaussian approximation was shown to be valid in Sec.~\ref{sec:gaussian}.
We use the same expansion through $O[(x-x')^4]$ of both $U(x,x')$ and the conformal factor $\Omega^2(x')$, described in the general discussion on Schwarzschild of this subsection.

Using this approach we have computed several components of the noise kernel when $\kappa \Delta t \ll 1$ and $\kappa \mathbf{r} \ll 1$
 through $O[(x-x')^{-4}]$.  The resulting expressions for an arbitrary separation of the points are too long to display in full here. If the points are separated along
the time direction we get the following result for the $N_{t~t'~}^{~t~t'}$ component:
\bea N_{t~t'~}^{~t~t'} &=&  \frac{1}{4 \pi^4 f^4} \left[ \frac{1}{\Delta t^8} - \frac{(1-f)^2}{72 r^2 \Delta t^6}
   +\frac{(1-f)^2 (1-2f+3f^2)}{864 r^4 \Delta t^4} \right. \nonumber \\
                                         & & \nonumber \\
                                         & & \left. - 5 \kappa^2 \left( \frac{1}{18 \Delta t^6} + \frac{(1-f)^2}{864 r^2 \Delta t^4} \right)
   + \kappa^4 \frac{17}{270 \Delta t^4} \right]  \,,
\label{Ntttt3}
\eea
which agrees with the expansion of Eq.~\eqref{kappa-r-large} through $O[(\kappa \Delta t)^{-4}]$.
If the points are separated along the radial direction, then we find
\bea N_{t~t'~}^{~t~t'} &=& \frac{1}{2 \pi^4} \left[ \frac{f^4}{2 \Delta r^8} - \frac{(1-f)f^3}{r \Delta r^7}
   + \frac{(1-f)(89-169f)f^2}{144 r^2 \Delta r^6} \right. \nonumber \\
                                         & & \nonumber \\
                                         & & \left. + \frac{(1-f)(1577 - 292 f - 441 f^2) f}{432 r^3 \Delta r^5} \right. \nonumber \\
                                         & & \nonumber \\
                                         & & \left. - \frac{(1-f)(240199 - 383185 f + 98655 f^2 + 18315 f^3)}{25920 r^4 \Delta r^4} \right. \nonumber \\
                                         & & \nonumber \\
                                         & & \left. + \kappa^2 \left( \frac{f^2}{36 \Delta r^6} - \frac{(1-f)f}{36 r \Delta r^5}
   + \frac{(1-f)(7-39f)}{1728 r^2 \Delta r^4} \right) \right. \nonumber \\
                                         & & \nonumber \\
                                         & & \left. - \frac{\kappa^4}{270 \Delta r^4} \right] .
\label{Ntttt4}
\eea
Note that the limit $\kappa \rightarrow 0$ of Eqs.~\eqref{Ntttt3}-\eqref{Ntttt4}
corresponds to the Boulware vacuum and the limit $f \rightarrow 1$ corresponds
to the flat space limit.
The latter coincides through $O[(\kappa\Delta t)^{-4}]$ and $O[(\kappa\Delta r)^{-4}]$, respectively, with the exact result in Eq.~\eqref{NKhfs} for the appropriate splitting of the points. Moreover, this coincidence is exact for zero temperature.

As discussed in Sec.~\ref{sec:general}, the noise kernel has two properties which can be used to check our calculations.
One of these, given in Eq.~\eqref{conservation}, is that the noise kernel should
be separately conserved at the points $x$ and $x'$.
The other, given in Eq.~\eqref{partial-traces}, is the vanishing of the partial
traces.
Enough components have been computed when $\kappa |\Delta t| \ll 1$ and $\kappa \mathbf{r} \ll 1$  that we have been able to check
all of the partial traces and all of the conservation conditions which involve the component $N_{t~t'~}^{~t~t'}$.  In each case they
 are satisfied to the order to which our computations are valid: $O[(x-x')^{-4}]$ for the partial traces and $O[(x-x')^{-5}]$ for the
 conservation conditions.

\section{Discussion}
\label{sec:discussion}

Using Page's approximation for the Euclidean Green function of a conformally invariant scalar field in the optical Schwarzschild spacetime, which is conformal to the static region of Schwarzschild spacetime, we have computed an expression for the Wightman function when the field is in a thermal state at an arbitrary temperature.  For the case that the temperature
 is equal to $(8 \pi M)^{-1}$ and one conformally transforms to Schwarzschild spacetime this corresponds to the Hartle-Hawking state.  This expression is exact for flat space and is valid through order $(x-x')^2$ in the optical Schwarzschild spacetime. From this expression for the Wightman function we have calculated the exact noise kernel in flat space and several components of an approximate one in Schwarzschild spacetime.  The latter is obtained by conformally transforming the noise kernel in the optical Schwarzschild spacetime to Schwarzschild spacetime.  We have shown that, unlike for the case of the stress tensor expectation value, this transformation is trivial.  In both the flat space and Schwarzschild cases we have restricted our attention to point separations which are either spacelike or timelike and we do not consider the limit in which the points come together.

For Schwarzschild spacetime we have considered two separate but related approximations for the noise kernel. The first one is valid for small separations (compared to the typical curvature radius scale) and arbitrary temperature. In this case we have explicitly computed one component of the noise kernel.
Note that although the Hartle-Hawking state corresponds to a specific temperature, given by Eqs.~(\ref{eq:temperature})-(\ref{eq:surf_grav}), our results also apply to any other temperature since we kept $\kappa$ arbitrary in all our expressions. The states for those other values of the temperature are singular on the horizon (e.g.\ the expectation value of the stress tensor diverges there), but can sometimes be of interest (e.g.\ the Boulware vacuum corresponds to the particular case of $T=0$).
The second approximation corresponds to further restricting to low enough temperatures or points that are close enough together so that the separation is much smaller than the inverse temperature. We have computed several components of the noise kernel within this approximation.

The component $N_{t~t'~}^{~t~t'}$ is displayed for both flat space~\eqref{NKhfs} and Schwarzschild spacetime.  In Schwarzschild spacetime it has been computed when the point separation is only in the time direction and the product of the temperature and the point separation is not assumed to be small~\eqref{kappa-r-large}. It has also been computed for an arbitrary spacelike or timelike separation of the points when the product of the temperature and point separation is small.  In this case, because of its length the expression is shown only for a point separation purely in the time direction~\eqref{Ntttt3} and for a point separation purely in the radial direction~\eqref{Ntttt4}.

 We have performed several nontrivial checks to verify our results. In both the
hot flat space case and in Schwarzschild spacetime we have checked both the
conservation and partial trace properties given in Eqs.~\eqref{conservation}
and~\eqref{partial-traces}.  For hot flat space
 these properties are satisfied exactly.  For Schwarzschild spacetime, where our
expression for the noise kernel is approximate, we have shown that the relevant
quantities vanish
 up to the order to which the approximation is valid. Furthermore, as an
additional check of the
 result~\eqref{kappa-r-large} for Schwarzschild spacetime when the separation is
in the time direction and the product of the temperature and the time separation
is not assumed to be small, we have considered two different ways of obtaining
the flat space limit of our result. Firstly, one can compare
with the hot flat space result~\eqref{NKhfs} by taking the limit $M \rightarrow
0$. Secondly, one can compare with Eq.~\eqref{N-Rindler} for the Minkowski
vacuum in Rindler coordinates by taking the limit $M \rightarrow \infty$ near
the horizon.

There are several more or less immediate generalizations of our work.
First, although the noise kernel corresponds to the expectation value of the anticommutator of the stress tensor, our results are also valid for other orderings of the stress tensor operator (in fact for any 2-point function of the stress tensor). That is always true for spacelike separated points because the commutator of any local operator (such as the stress tensor) vanishes as a consequence of the microcausality condition. Moreover, since for conformal fields in Schwarzschild the commutator of the field, $iG(x,x')$, also vanishes for timelike separated points up to the order to which we are working, the previous statement also holds for timelike separations in our case.\footnote{In general one would need to use the appropriate prescription when analytically continuing the Euclidean Green function to obtain the Wightman function for timelike separated points in the Lorentzian case, and use expressions analogous to Eqs.~(\ref{eq:temperature})-(\ref{eq:surf_grav}) but without symmetrizing with respect to the two points. One can explicitly see how this is done in Ref.~\cite{perez-nadal10}.}
Second, since the Gaussian approximation is valid for any ultra-static spacetime which is conformal to an Einstein metric (a solution of the Einstein equation in vacuum, with or without cosmological constant) \cite{page}, one can straightforwardly extend our calculation to all those cases by taking the general expression for the Wightman function under the Gaussian approximation, given by Eq.~(\ref{Gwgauss}), and substituting it into the general expression for the noise kernel given in Eqs.~\eqref{general-noise-kernel} and~\eqref{generalnoise}.

One of the most interesting uses of the noise kernel is to investigate the
effects of quantum fluctuations near the horizon of the black hole. For instance, there have been claims in the literature that the size of the horizon could exhibit fluctuations induced by the vacuum fluctuations of the matter fields which are much larger than the Planck scale (even for relatively short timescales of the order of the Schwarzschild radius, i.e. much shorter than the evaporation time) \cite{sorkin95,sorkin97,casher97,marolf03}. So far all these studies have been based on semi-qualitative arguments. However, one should in principle be able to address this issue by computing the quantum correlation function of the metric perturbations, including the effects of loops of matter fields, with the method outlined in the introduction.
As a matter of fact, part of the information on the corresponding induced curvature fluctuations is already directly available from our results. Indeed, at one loop the correlator of the Ricci tensor (or, equivalently, the Einstein tensor) is gauge invariant\footnote{This quantity is gauge invariant because the Ricci tensor of the Schwarzschild background vanishes, as does its Lie derivative with respect to an arbitrary vector.} and it is immediately given by the stress tensor correlator \cite{perez-nadal10}. Unlike the Riemann tensor, the Ricci tensor does not entirely characterize the local geometry. In order to get the full information about the quantum fluctuations of the geometry at this order, one needs to use Eq.~\eqref{h-h-s} or a related one. In that case, the noise kernel for arbitrary pairs of points is a crucial ingredient. Strictly speaking it is important that the noise kernel, although divergent in the coincidence limit, is a well-defined distribution. Our result for separate points does not completely characterize such a distribution since it does not specify the appropriate integration prescription in the coincidence limit. This can, nevertheless, be obtained using the method in Appendix~C of Ref.~\cite{HuRou07} (see also Ref.~\cite{froeb11a} for cosmological examples).

It is worthwhile to discuss briefly how the present paper is related to an earlier study of the noise kernel in Schwarzschild spacetime~\cite{PH03}, which also considered a conformal scalar field and made use of Page's Gaussian approximation. The main interest there was evaluating the noise kernel in the coincidence limit. In order to get a finite result, the Hadamard elementary solution was subtracted from the Wightman function before evaluating the noise kernel.
Since the Hadamard elementary solution coincides with the $\kappa = 0$ expression for the Gaussian approximation through order $(x-x')^2$, which is the order through which the approximation is valid for the optical Schwarzschild spacetime, their subtracted Wightman function will also be valid through that order.
The fact that they found a non-vanishing trace for their noise kernel is also compatible with our results because, as we have reasoned, the noise kernel should only be valid through order $(x-x')^{-4}$ when the Gaussian approximation for the Wightman function is employed (and through order $(x-x')^{-2}$ when using the subtracted Wightman function, whose leading term is $O(1)$ rather than $O[(x-x')^{-2}]\,$).
Instead, one would need an expression for the noise kernel accurate through order $(x-x')^0$ or higher to get a vanishing trace in the coincidence limit.
In contrast, for the reasons given in the introduction, here we consider the unsubtracted noise kernel, which is indispensable to obtain the quantum correlation function for the metric perturbations as the subtracted one would lead for instance to a vanishing result --and no fluctuations-- for the Minkowski vacuum. Furthermore, in this way one can still get useful and accurate information for the terms of order $(x-x')^{-8}$ through $(x-x')^{-4}$, which dominate at small separations.

\acknowledgments
This paper is based on a chapter of the PhD thesis of the first author~\cite{thesis}.
A.~E.\ thanks Jos\'e~M.\ Mart\'{\i}n-Garc\'{\i}a for help with the \texttt{xAct} package for
tensor computer algebra~\cite{xact} which was employed in a related calculation.
This work was supported in part by the National Science Foundation under grants PHY-0601550 and
PHY-0801368 to the University of Maryland, and grant
PHY-0856050 to Wake Forest University.

\appendix*

\section{Noise kernel and conformal transformations}
\label{app:conformal}

In this appendix we derive the result for the rescaling of the noise kernel under conformal transformations. We provide two alternative proofs based respectively on the use of quantum operators and on functional methods.

First, we start by showing how the classical stress tensor of a conformally invariant scalar field rescales under a conformal transformation $g_{ab} \to \tilde{g}_{ab} = \Omega^2(x)\, g_{ab}$. The key point is that the classical action of the field, $S[\phi,g]$, remains invariant (up to surface terms) if one rescales appropriately the field: $\phi \to \tilde{\phi} = \Omega^{(2-D)/2} \phi$. Taking that into account, one easily gets the result from the definition of the stress tensor as a functional derivative of the classical action:
\begin{equation}
\tilde{T}_{ab}
=  \frac{2\tilde{g}_{ac} \tilde{g}_{bd}}{\sqrt{-\tilde{g}}}
\frac{\delta S[\tilde{\phi},\tilde{g}]}{\delta\tilde{g}_{cd}}
=  \frac{2\tilde{g}_{ac} \tilde{g}_{bd}}{\sqrt{-\tilde{g}}}
\frac{\delta S[\phi,g]}{\delta\tilde{g}_{cd}}
= \Omega^{2-D} \frac{2g_{ac} g_{bd}}{\sqrt{-g}}
\frac{\delta S[\phi,g]}{\delta g_{cd}}
= \Omega^{2-D}\,  T_{ab}
\label{eq:rescaling2} .
\end{equation}

\subsection{Proof based on quantum operators}

A possible way of proving Eq.~(\ref{eq:rescaling1}) is by promoting  the classical field $\phi$ in Eq.~(\ref{eq:rescaling2}) to an operator in the Heisenberg picture. The operator $\hat{T}_{ab}(x) $ would be divergent because it involves products of the field operator at the same point. However, in order to calculate the noise kernel what one actually needs to consider is $\hat{t}_{ab} (x) = \hat{T}_{ab} (x) - \langle \hat{T}_{ab} (x) \rangle$ and this object is UV finite, i.e., its matrix elements $\langle\Phi| \hat{t}_{ab} (x)| \Psi\rangle$ for two arbitrary states $| \Psi\rangle$ and $| \Phi\rangle$ (not necessarily orthogonal) are UV finite because Wald's axioms \cite{bir-dav-book} guarantee that $\langle\Phi| \hat{T}_{ab} (x)| \Psi\rangle$ and $\langle\Phi | \Psi\rangle \langle \hat{T}_{ab} (x) \rangle $ have the same UV divergences and they cancel out.
Therefore, one can proceed as follows. One starts by introducing a UV regulator (it is useful to consider dimensional regularization since it is compatible with the conformal symmetry for scalar and fermionic fields, but this is not indispensable since we will remove the regulator at the end without having performed any subtraction of non-invariant counterterms). One can next apply the operator version of Eq.~(\ref{eq:rescaling2}) to the operators $\hat{t}_{ab} (x)$ appearing in Eq.~(\ref{NKdef}) defining the noise kernel. Since all UV divergences cancel out, as argued above, we can then safely remove the regulator and are finally left with Eq.~(\ref{eq:rescaling1}).

\subsection{Proof based on functional methods}

An alternative way of proving Eq.~(\ref{eq:rescaling1}) is by analyzing how the closed-time-path (CTP) effective action $\Gamma[g,g']$ changes under conformal transformations. This effective action results from treating $g_{ab}$ and $g'_{ab}$ as external background metrics and integrating out the quantum scalar field within the CTP formalism \cite{chou85}:
\begin{equation}
e^{\imath \Gamma[g,g']} =
\int \mathcal{D}\varphi_\mathrm{f}\,  \mathcal{D}\varphi_\mathrm{i}\,
\mathcal{D}\varphi'_\mathrm{i}\,
\rho[\varphi_\mathrm{i},\varphi'_\mathrm{i}]
\int^{\varphi_\mathrm{f}}_{\varphi_\mathrm{i}} \mathcal{D}\phi \,
e^{\imath S_\mathrm{g}[g] + \imath S[\phi,g]}
\int^{\varphi_\mathrm{f}}_{\varphi'_\mathrm{i}} \mathcal{D}\phi' \,
e^{-\imath S_\mathrm{g}[g'] - \imath S[\phi',g']}
\label{eq:ctp1} ,
\end{equation}
where $\rho[\varphi_\mathrm{i},\varphi'_\mathrm{i}]$ is the density matrix functional for the initial state of the field
\footnote{Under appropriate conditions it is also possible to consider asymptotic initial states. For instance, given a static spacetime, a generalization to the CTP case of the usual $\imath\epsilon$ prescription involving a small Wick rotation in time selects the ground state of the Hamiltonian associated with the time-translation invariance as the initial state.}
(in particular one has $\rho[\varphi_\mathrm{i},\varphi'_\mathrm{i}] = \Psi[\varphi_\mathrm{i}] \Psi^*[\varphi'_\mathrm{i}]$ for a pure initial state with wave functional $\Psi[\varphi_\mathrm{i}] = \langle \varphi_\mathrm{i} | \Psi \rangle$ in the Schr\"odinger picture),
$S_\mathrm{g}[g]$ is the gravitational action including local counterterms, $S[\phi,g]$ is the action for the scalar field, and the two background metrics are also taken to coincide at the same final time at which the final configuration of the scalar field for the two branches are identified and integrated over.
The fields $\varphi_\mathrm{f}$ on the one hand and $\{\varphi_\mathrm{i},\varphi'_\mathrm{i}\}$ on the other, correspond to the values of the field restricted respectively to the final and initial Cauchy surfaces, and their functional integrals are over all possible configurations of the field on those surfaces.
Integrating out the scalar field gives rise to UV divergences, but they can be dealt with by renormalizing the cosmological constant and the gravitational coupling constant as well as introducing local counterterms quadratic in the curvature in the bare gravitational action $S_\mathrm{g}[g]$, so that the total CTP effective action is finite. After functionally differentiating and identifying the two background metrics, one gets the renormalized expectation value of the stress tensor operator together with the contributions from the gravitational action \cite{martin99b, roura08_appC}:
\begin{equation}
g_{ac} \, g_{bd} \, \frac{2}{\sqrt{-g}}
\left. \frac{\delta \Gamma[g,g']}{\delta g_{cd}} \right|_{g'=g}
\! = - \frac{1}{8\pi G} \big( G_{ab} + \Lambda g_{ab} \big)
+ \big\langle \hat{T}_{ab} \big\rangle_\mathrm{ren}
\label{eq:Tab} ,
\end{equation}
where the contribution from the counterterms quadratic in the curvature has been absorbed in $\langle \hat{T}_{ab} \rangle_\mathrm{ren}$. The renormalized gravitational coupling and consmological constants, $G$ and $\Lambda$, depend on the renormalization scale, but the expectation value also depends on it in such a way that the total expression is renormalization-group invariant since that is the case for the effective action. The equation that one obtains by equating the right-hand side of Eq.~(\ref{eq:Tab}) to zero governs the dynamics of the mean field geometry in semiclassical gravity, including the back-reaction effects of the quantum matter fields.

On the other hand, the noise kernel can be obtained by functionally differentiating twice the imaginary part of the CTP effective action:
\begin{equation}
N_{abc'd'}(x,x') =
g_{ae}(x) g_{bf}(x) g_{c'g'}(x') g_{d'h'}(x') \, \frac{4}{\sqrt{g(x)g(x')}}
\left. \frac{\delta^2\, \mathrm{Im}\, \Gamma[g,g']}{\delta g_{e'f'}(x) \delta g_{g'h'}(x')} \right|_{g'=g}
\label{eq:noiseA1} .
\end{equation}
It is well-known that the imaginary part of the effective action does not contribute to the equations of motion for expectation values derived within the CTP formalism, like Eq.~(\ref{eq:Tab}), which are real. Furthermore, one can easily see from Eq.~(\ref{eq:ctp1}) that, since it is real, the gravitational action (whose contribution can be factored out of the path integral) only contributes to the real part of the effective action. In particular this means that the counterterms and the renormalization process have no effect on the noise kernel, which will be a key observation in order to prove Eq.~(\ref{eq:rescaling1}). Indeed, let us start with Eq.~(\ref{eq:ctp1}) for the conformally related metric and scalar field, $\tilde{g}_{ab}$ and $\tilde{\phi}$, and assume that we use dimensional regularization:
\begin{eqnarray}
e^{\imath \Gamma[\tilde{g},\tilde{g}']} &=&
e^{\imath S_\mathrm{g}[\tilde{g}] - \imath S_\mathrm{g}[\tilde{g}']}
\int \mathcal{D}\tilde{\varphi}_\mathrm{f}\,  \mathcal{D}\tilde{\varphi}_\mathrm{i}\,
\mathcal{D}\tilde{\varphi}'_\mathrm{i}\,
\tilde{\rho}[\tilde{\varphi}_\mathrm{i},\tilde{\varphi}'_\mathrm{i}]
\int^{\tilde{\varphi}_\mathrm{f}}_{\tilde{\varphi}_\mathrm{i}} \mathcal{D}\tilde{\phi} \,
e^{\imath S[\tilde{\phi},\tilde{g}]}
\int^{\tilde{\varphi}_\mathrm{f}}_{\tilde{\varphi}'_\mathrm{i}} \mathcal{D}\tilde{\phi}' \,
e^{- \imath S[\tilde{\phi}',\tilde{g}']} \nonumber \\
&=& e^{\imath (S_\mathrm{g}[\tilde{g}] - S_\mathrm{g}[g])
- \imath(S_\mathrm{g}[\tilde{g}'] - S_\mathrm{g}[g'])}
\int \mathcal{D}\varphi_\mathrm{f}\,  \mathcal{D}\varphi_\mathrm{i}\,
\mathcal{D}\varphi'_\mathrm{i}\,
\rho[\varphi_\mathrm{i},\varphi'_\mathrm{i}]
\int^{\varphi_\mathrm{f}}_{\varphi_\mathrm{i}} \mathcal{D}\phi \,
\left| \frac{\mathcal{D}\tilde{\phi}}{\mathcal{D}\phi} \right|
e^{\imath S_\mathrm{g}[g] + \imath S[\phi,g]} \nonumber \\
&&\times \int^{\varphi_\mathrm{f}}_{\varphi'_\mathrm{i}} \mathcal{D}\phi' \,
\left| \frac{\mathcal{D}\tilde{\phi}'}{\mathcal{D}\phi'} \right|
e^{-\imath S_\mathrm{g}[g'] - \imath S[\phi',g']}
\label{eq:ctp2} ,
\end{eqnarray}
where we have taken into account in the second equality the fact that dimensional regularization is compatible with the invariance of the classical action $S[\tilde{\phi},\tilde{g}]$ under conformal transformations (since it is invariant in arbitrary dimensions). We also considered that the initial states of the scalar field are related by
\begin{eqnarray}
\tilde{\rho}\big[\tilde{\varphi}_\mathrm{i}(x),\tilde{\varphi}'_\mathrm{i}(x')\big]
&=& \Omega^{(D-2)/4}_\mathrm{i}(x) \, \Omega^{(D-2)/4}_\mathrm{i}(x')\,\,
\rho[\varphi_\mathrm{i}(x),\varphi'_\mathrm{i}(x')] \nonumber \\
&=& \Omega^{(D-2)/4}_\mathrm{i}(x) \, \Omega^{(D-2)/4}_\mathrm{i}(x')\,\,
\rho \Big[ \Omega^{-1}_\mathrm{i}(x)\tilde{\varphi}_\mathrm{i}(x),
\Omega^{-1}_\mathrm{i}(x)\tilde{\varphi}'_\mathrm{i}(x') \Big]
\label{eq:state} ,
\end{eqnarray}
where the conformal factor $\Omega^2_\mathrm{i}$ is  restricted to the initial Cauchy surface and so are the points $\{x,x'\}$ in this equation.
(This relation between the initial states is the choice compatible with conformal invariance after one takes into account the relation between $\tilde{\phi}$ and $\phi$, and the prefactor is determined by requiring that the state remains normalized.)
The logarithm of the functional Jacobian $|\mathcal{D}\tilde{\phi}/\mathcal{D}\phi|$ is divergent but formally zero in dimensional regularization%
\footnote{This can be seen by taking Eq.~(18) in Ref.~\cite{fujikawa80} and using dimensional regularization \cite{bir-dav-book} to evaluate the trace of the heat kernel appearing there.
Any possible dependence left on the conformal factor evaluated at the initial
or final Cauchy surfaces would correspond to a prefactor on the right-hand
side of Eq.~(\ref{eq:ctp2}), and would not contribute to the noise kernel (or the expectation value of the stress tensor) at any intermediate time since it involves
functionally differentiating the logarithm of that expression with respect to the
metric at such intermediate times.},
so that we can take $|\mathcal{D}\tilde{\phi}/\mathcal{D}\phi| = 1$ in both path integrals on the right-hand side of Eq.~(\ref{eq:ctp2}). Taking all this into account, we are left with
\begin{equation}
\Gamma[\tilde{g},\tilde{g}']
= \Gamma[g,g'] + (S_\mathrm{g}[\tilde{g}] - S_\mathrm{g}[g])
- (S_\mathrm{g}[\tilde{g}'] - S_\mathrm{g}[g'])
\label{eq:ctp3} ,
\end{equation}
where the last two pairs of terms on the right-hand side correspond to the difference between the bare gravitational actions of the two conformally related metrics in dimensional regularization; note that whereas each bare action is separately divergent, the difference $S_\mathrm{g}[\tilde{g}] - S_\mathrm{g}[g]$ is finite. When working in dimensional regularization,
conformally invariant fields only exhibit divergences associated with counterterms quadratic in the curvature. These terms lead to the standard result for the trace anomaly of the stress tensor when one takes the functional derivative of Eq.~(\ref{eq:ctp3}) with respect to the conformal factor, which can be shown to be equivalent to the trace of Eq.~(\ref{eq:Tab}).

The key aspect for our purposes is that the extra terms on the right-hand side of Eq.~(\ref{eq:ctp3}) only change the real part of the CTP effective action, as already mentioned above, so that the imaginary part remains invariant under conformal transformations. Starting with Eq.~(\ref{eq:noiseA1}) for the metric $\tilde{g}_{ab}$ and taking into account the invariance of the imaginary part of the CTP effective action under conformal transformations, one immediately obtains
\begin{equation}
\tilde{N}_{abc'd'}(x,x') =  \Omega^{2-D}(x)\, \Omega^{2-D}(x')\, N_{abc'd'}(x,x') ,
\end{equation}
in agreement with Eq.~(\ref{eq:rescaling1}).
Note that we have employed dimensional regularization in our argument for simplicity, but one would reach the same conclusion if other regularization schemes had been used. In those cases one would get in general a contribution to the analog of Eq.~(\ref{eq:ctp3}) from the change of the functional measure, but it would only affect the real part of the effective action
(see Ref.~\cite{fujikawa80}, where the calculations are performed in Euclidean time, and analytically continue the result to Lorentzian time)
and one could still apply exactly the same argument as before.


\begin{thebibliography}{100}

\bibitem{page} D. N. Page,
Phys. Rev. D {\bf 25}, 1499 (1982).

\bibitem{Ford82}
L.H. Ford, Ann. Phys. (N.Y.) {\bf 144}, 238 (1982).

\bibitem{Hu89}
B.-L. Hu.
{\em Physica} A {\bf 158}, 399 (1989).

\bibitem{kuo-ford}
C.-I. Kuo and L.H. Ford,
Phys. Rev. D {\bf 47}, 4510 (1993).

\bibitem{wu-ford-1}
C.-H. Wu and L.H. Ford,
Phys. Rev. D {\bf 60}, 104013 (1999).

\bibitem{wu-ford-2}
C.-H. Wu and L.H. Ford, Phys. Rev. D {\bf 64}, 045010 (2001)

\bibitem{PH97} N. G. Phillips and B. L. Hu, Phys. Rev. D
{\bf 55}, 6132 (1997).

\bibitem{PH00}
N. G.  Phillips and  B. L. Hu, Phys. Rev. D {\bf 62}, 084017 (2000)

\bibitem{PH00-2}
B.L. Hu and N.G. Phillips,
Int. J. Theor. Phys. {\bf 39}, 1817 (2000).


\bibitem{semiclassical gravity}
Ya. Zel'dovich and A. Starobinsky, Zh. Eksp. Teor. Fiz {\bf 61}, 2161
(1971) [Sov. Phys.- JETP {\bf 34}, 1159 (1971)]
L. Grishchuk, Ann. N. Y. Acad. Sci. 302, 439 (1976).
B. L. Hu and L. Parker, Phys. Lett. 63A, 217 (1977).
B. L. Hu  and L. Parker, Phys. Rev. {\bf D17}, 933 (1978).
F. V. Fischetti, J. B. Hartle and B. L. Hu, Phys. Rev. {\bf D20},
1757 (1979).
J. B. Hartle and B. L. Hu, Phys. Rev. {\bf D20}, 1772 (1979). {\bf
21}, 2756 (1980).
J. B. Hartle, Phys. Rev. D23, 2121 (1981). P. A. Anderson, Phys. Rev.
D28, 271 (1983); D29, 615 (1984). E. Calzetta and B. L. Hu, Phys. Rev. D {\bf 35}, 495 (1987).
 A. Campos and E. Verdaguer, Phys. Rev. D {\bf 49}, 1861 (1994).

\bibitem{And-Mol-Mot-1} P. R. Anderson, C. Molina-Paris, and E. Mottola, Phys. Rev. D {\bf 67},
024026 (2003).

\bibitem{HRV04} B. L. Hu, Albert Roura, and Enric Verdaguer,
Phys. Rev. D {\bf 70}, 044002 (2004).

\bibitem{perez-nadal08}
G. P\'erez-Nadal, A. Roura, and E. Verdaguer,
Phys. Rev. D {\bf 77}, 124033 (2008);
Class. Quant. Grav. {\bf 25}, 154013 (2008).

\bibitem{And-Mol-Mot-2} P. R. Anderson, C. Molina-Paris, and E. Mottola,
Phys. Rev. D {\bf 80}, 084005 (2009).

\bibitem{froeb11b}
M. Fr\"oeb, D. B. Papadopoulos, A. Roura, and E. Verdaguer,
``Nonperturbative semiclassical stability of de Sitter spacetime"
(in preparation).

\bibitem{ford-structure-1} C.-H. Wu, K.-W. Ng, and L. H. Ford,
Phys. Rev. D 75, 103502 (2007).

\bibitem{roura08}
A. Roura and E. Verdaguer,
Phys. Rev. D {\bf 78}, 064010 (2008).

\bibitem{perez-nadal10}
G. P\'erez-Nadal, A. Roura, and E. Verdaguer,
JCAP {\bf 05} (2010) 036.

\bibitem{ford-structure-2} L. H. Ford, S. P. Miao, K.-W. Ng, R. P. Woodard, and C.-H. Wu,
Phys. Rev. D {\bf 82}, 043501 (2010).

\bibitem{froeb11a}
M. Fr\"oeb, A. Roura, and E. Verdaguer,
``One-loop gravitational wave spectrum in de Sitter spacetime"
(in preparation).

\bibitem{HuRou07} B. L. Hu  and Albert Roura,
Phys. Rev. D {\bf 76}, 124018 (2007).

\bibitem{HuRou07b} B. L. Hu  and Albert Roura,
Int. J. Theor. Phys.  {\bf 46}, 2204 (2007).

\bibitem{HuKinTh} B. L. Hu, Int. J. Theor. Phys. {\bf 41}, 2111 (2002)[[gr-qc/0204069].

\bibitem{Hu99} B. L. Hu,  Int. J. Theor. Phys. {\bf 38}, 2987 (1999) [gr-qc/9902064].

\bibitem{martin99a} R. Mart\'\i n and E. Verdaguer,
Phys. Lett. B {\bf 113}, 465 (1999).

\bibitem{martin99b} R. Mart\'\i n and E. Verdaguer,
Phys. Rev. D {\bf 60}, 084008 (1999).

\bibitem{martin00} R. Mart\'\i n and E. Verdaguer,
Phys. Rev. D {\bf 61}, 124024 (2000).

\bibitem{stograLivRev} B. L. Hu and E. Verdaguer,
Liv. Rev. Rel. {\bf 11}, 3 (2008).

\bibitem{stograCQG}B. L. Hu and E. Verdaguer,
Class. Quant. Grav.  {\bf 20}, R1-R42 (2003).

\bibitem{ELE} E. Calzetta and B. L. Hu,    Phys. Rev. D {\bf 49}, 6636 (1994);
B. L. Hu and A. Matacz,    Phys. Rev. D {\bf 51}, 1577 (1995);  B. L.
Hu and S. Sinha,       Phys.  Rev. D {\bf 51}, 1587 (1995);
  A. Campos and E. Verdaguer,   Phys. Rev. D {\bf 53}, 1927 (1996);
  F. C. Lombardo and F. D. Mazzitelli,  Phys. Rev. D {\bf 55}, 3889 (1997).
  A. Campos and B. L. Hu, Phys. Rev. D {\bf 58} (1998) 125021

\bibitem{birrell94}
N.~D. Birrell and P.~C.~W. Davies,
{\em Quantum fields in curved space}
(Cambridge University Press, Cambridge, 1994).

\bibitem{CRVopensys} E. Calzetta, A. Roura, and E. Verdaguer,
Physica A {\bf 319}, 188 (2003).


\bibitem{HRS} B. L. Hu, A. Raval and S. Sinha,
``Notes on Black Hole Fluctuations  and Backreaction",
in {\it Black Holes, Gravitational Radiation and the Universe: Essays in honor
of C. V. Vishveshwara}, edited by B. Iyer and B. Bhawal (Kluwer Academic
Publishers, Dordrecht, 1998),
\eprint{gr-qc/9901010}.

\bibitem{SRH} S. Sinha, A. Raval and B. L. Hu, ``Black Hole Fluctuations
and Backreaction in Stochastic Gravity",  in Foundations of Physics
33 (2003) 37-64  [gr-qc/0210013]

\bibitem {york85}
J. W. York, Jr., Phys. Rev. D {\bf 31}, 775 (1985).

\bibitem {York}
J. W. York, Jr.,
Phys. Rev. D {\bf 28}, 2929 (1983);
\emph{ibid.} {\bf 33}, 2092 (1986).

\bibitem{York2} D. Hochberg and T. W. Kephart, Phys. Rev. D {\bf 47}, 1465 (1993); D. Hochberg, T. W. Kephart, and J. W. York, Jr., Phys. Rev. D {\bf 48}, 479 (1993);  P. R. Anderson, W. A. Hiscock, J. Whitesell, and J. W. York, Jr., Phys. Rev. D {\bf 50}, 6427 (1994).

\bibitem{burgess04}
C.~P. Burgess,
Living Rev. Rel. {\bf 7}, 5 (2004).

\bibitem{roura99b}
A. Roura and E. Verdaguer,
Int. J. Theor. Phys. {\bf 38}, 3123 (1999).

\bibitem{PH01}
N. G.  Phillips and  B. L. Hu, Phys. Rev. D {\bf 63}, 104001 (2001).

\bibitem{perez-nadal11}
G. P\'erez-Nadal, A. Roura, and E. Verdaguer,
``Stress tensor fluctuations in maximally symmetric spacetimes"
(in preparation).

\bibitem{park11}
S. Park, R. P. Woodard,
Phys. Rev. D {\bf 83}, 084049 (2011).

\bibitem{ChoHu noise kernelAdS} H. T. Cho and B. L. Hu, ``Stress-energy Tensor Correlators of a Quantum Field in Euclidean $R^N$ and $AdS^N$ spaces via the generalized zeta-function method."  Phys. Rev. D84, 044032 (2011) arXiv:1105.5308  II. Finite Temperature" (in preparation).      J. Physics (Conf. Ser.)   arXiv:1105.5302

\bibitem{PH03} N. G. Phillips  and B. L. Hu, Phys. Rev. D {\bf 67}, 104002  (2003).


\bibitem{christensen} S. M. Christensen,
Phys. Rev. D {\bf 14}, 2490 (1976).

\bibitem{MTW} C. W. Misner, K. S. Thorne, and J. A. Wheeler, {\it Gravitation} (Freeman, San Francisco, 1973).

\bibitem{bir-dav-book} See e.g. N. D. Birrell and P. C. W. Davies, {\it Quantum Fields in Curved Space} (Cambridge University Press, Cambridge, 1994).

\bibitem{aeh} See e.g. P. R. Anderson, A. Eftekharzadeh, B. L. Hu, Phys. Rev. D  {\bf 73}, 064023 (2006).

\bibitem{AndHu04} Paul R. Anderson and B. L. Hu, Phys. Rev. D 69, 064039 (2004)

\bibitem{gac} See e.g. Appendix D of P. B. Groves, P. R. Anderson, and E. D. Carlson, Phys. Rev. D {\bf 66}, 124017 (2002).

\bibitem{sorkin95}
R.~D. Sorkin,
in {\em Proceedings of the Conference on Heat Kernel Techniques
and Quantum Gravity}, edited by S.~A. Fulling,
Discourses in Mathematics and its Applications Vol. 4
(University of Texas Press, College Station, Texas, 1995),
\eprint{arXiv:gr-qc/9508002}.

\bibitem{sorkin97}
R.~D. Sorkin,
in {\em Proceedings of the First Australasian Conference on General
Relativity and Gravitation}, edited by D.~Wiltshire
(University of Adelaide, Adelaide, Australia, 1997),
\eprint{arXiv:gr-qc/9701056}.

\bibitem{casher97}
A.~Casher, F.~Englert, N.~Itzhaki, S.~Massar, and R.~Parentani.
{\em Nucl. Phys. B}, {\bf 484} 419 (1997).

\bibitem{marolf03}
D.~Marolf,
in {\em Particle physics and the universe},
edited by J.~Trampetic and J.~Wess
Springer Proc. Phys. Vol. 98 (Springer-Verlag, 2005),
\eprint{arXiv:hep-th/0312059}.

\bibitem{thesis} A. Eftekharzadeh, Ph.D. thesis, University of Maryland, 2007.

\bibitem{xact} J. M. Mart\'{\i}n-Garc\'{\i}a,
\url{http://metric.iem.csic.es/Martin-Garcia/xAct} .


\bibitem{chou85}
K. Chou, Z. Su, B. Hao, and L. Yu,
Phys. Rep. {\bf 118}, 1 (1985).

\bibitem{roura08_appC}
See Appendix~C in Ref.~\cite{roura08}.

\bibitem{fujikawa80}
K. Fujikawa,
Phys. Rev. Lett. {\bf 44}, 1733 (1980).


\end{thebibliography}
\end{document}